\documentclass[prl,aps,twocolumn]{revtex4}

\usepackage{graphicx}% Include figure files
\usepackage{dcolumn}% Align table columns on decimal point
\usepackage{bm}% bold math
\usepackage{amssymb}

\begin{document}

%\preprint{Preprint}

\title{Noncoplanar magnetic ordering driven by itinerant electrons on the pyrochlore lattice}

\author{Gia-Wei Chern}
\affiliation{Department of Physics, University of Wisconsin,
Madison, Wisconsin 53706, USA}

\date{\today}

\begin{abstract}
Exchange interaction tends to favor collinear or coplanar magnetic
orders in rotationally invariant spin systems. Indeed, such magnetic
structures are usually selected by thermal or quantum fluctuations
in highly frustrated magnets. Here we show that a complex {\em
noncoplanar} magnetic order with a quadrupled unit cell is
stabilized by itinerant electrons on the pyrochlore lattice.
Specifically we consider the Kondo-lattice and Hubbard models at
quarter filling. The electron Fermi `surface' at this filling factor
is topologically equivalent to three intersecting Fermi circles.
Perfect nesting of the Fermi lines leads to magnetic ordering with
multiple wavevectors and a definite handedness. The chiral order
might persist without magnetic order in a chiral spin liquid at
finite temperatures.
\end{abstract}

\maketitle

\begin{figure} [t]
\includegraphics[width=0.86\columnwidth]{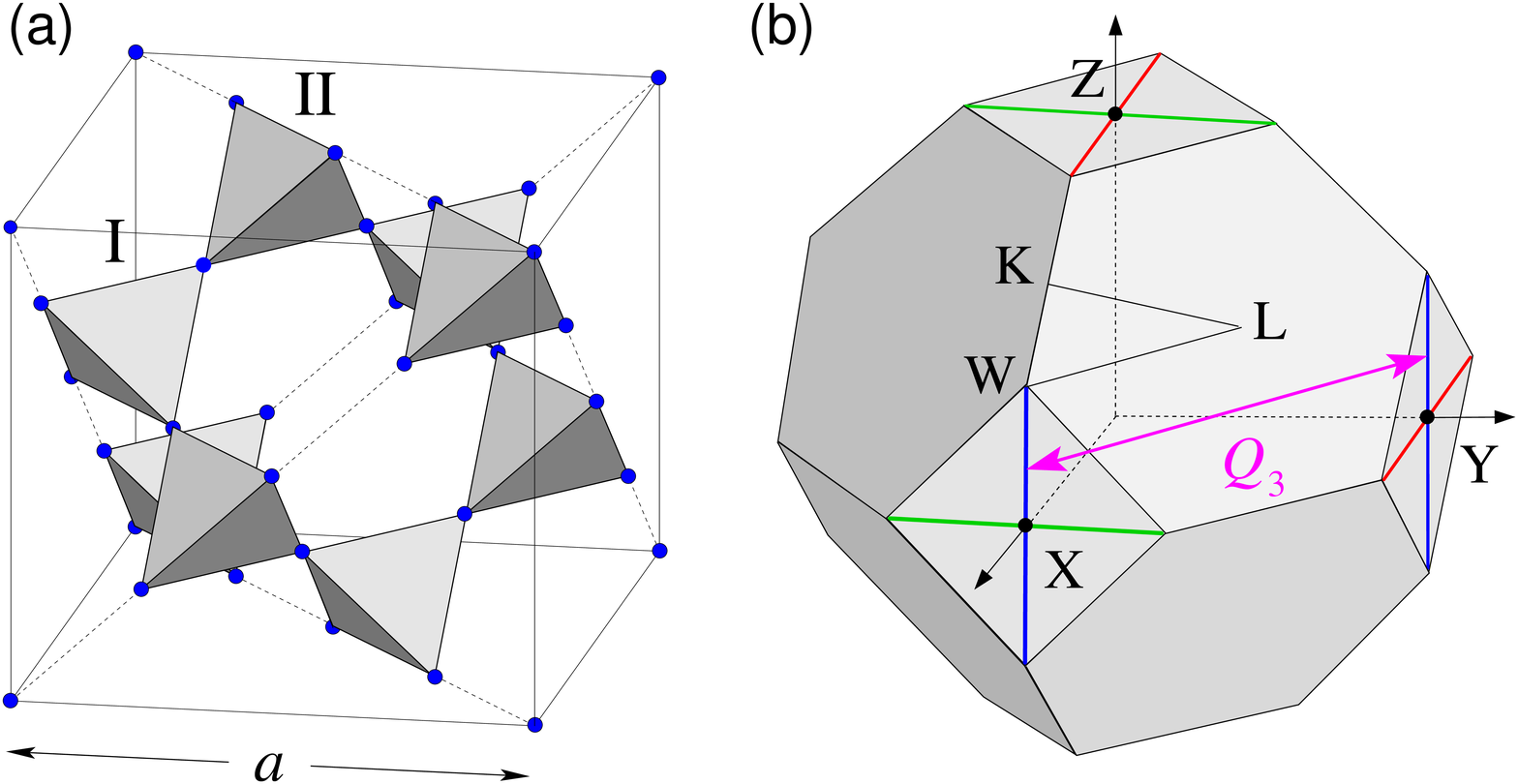}
\includegraphics[width=0.33\columnwidth]{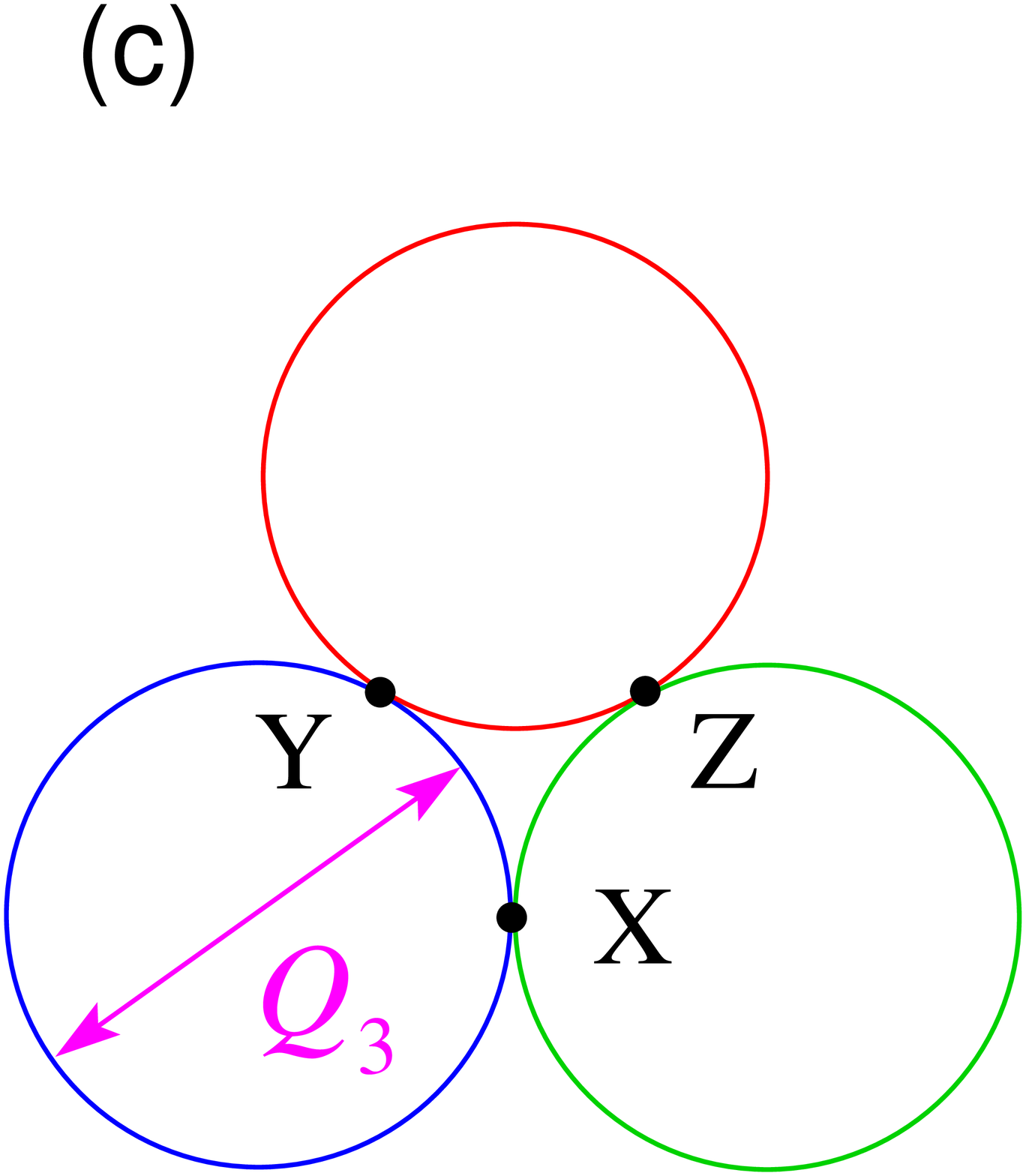}
\includegraphics[width=0.64\columnwidth]{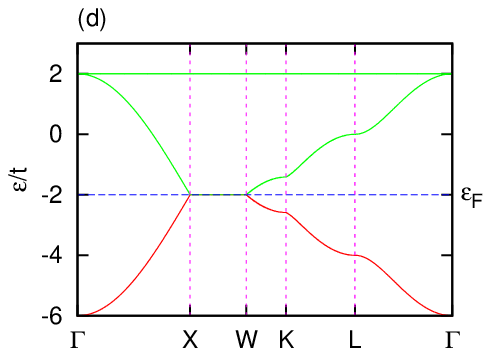}
\caption{\label{fig:spectrum} (a) A conventional cubic unit cell of
the pyrochlore lattice. I and II denote the two types of tetrahedra
with different orientations. (b) First Brillouin zone of the fcc
lattice. At quarter filling, the electron Fermi `surface' consists
of diagonal lines on the square face of the Brillouin zone. (c)
Topologically, these Fermi lines are equivalent to three closed
loops interceting at the X points of the Brillouin zone. (d) Band
structure of the tight-binding model on pyrochlore lattice. The
segment XW corresponds to one-quarter of a Fermi line at quarter
filling.}
\end{figure}

Magnets with geometrical frustration have fascinated physicists for
more than a decade as models of strongly interacting systems. A
prominent example of magnetic frustration in high dimensions is the
nearest-neighbor antiferromagnet on the {\em pyrochlore} lattice
shown in Fig.~\ref{fig:spectrum}(a). For classical Heisenberg spins,
strong geometrical frustration prevents the magnet from settling in
a long-range magnetic order even at zero temperature
\cite{moessner98}. Instead, the ground state retains a finite
zero-point entropy and is susceptible to small perturbations such as
anisotropies, further-neighbor couplings, and dipolar interactions.
In real compounds, magnetic frustration is often relieved when spins
couple to other degrees of freedom. An intensively studied case is
spin-lattice coupling in chromium spinels where the magnetic
transition is accompanied by a lattice distortion
\cite{lee00,ot02,chern06}. Another well known example is magnetic
phase transition faciliated by orbital order in frustrated
spin-orbital systems \cite{lee04,garlea08,ot04,chern09}.

Metallic pyrochlore magnets such as Kondo-lattice or double-exchange
models pose a different challenge for theorists due to the nonlocal
nature of the electron-mediated interactions. Indeed, despite
considerable effort \cite{ikoma03,ikeda08,motome10,lacroix}, a
complete picture of the phase diagram is still lacking. An early
study of the double-exchange model on pyrochlore lattice revealed a
rich mean-field phase diagram \cite{ikoma03}. However, the
calculation only considered magnetic orders which preserve the
lattice translational symmetry, hence precluding magnetic structures
with multiple wavevectors. A recent Monte Carlo study of the same
model provided an unbiased phase diagram at large Hund's coupling
\cite{motome10}. In particular, a peculiar state with electronic
phase separation was observed. Magnetic orders in the weak-coupling
regime remain unclear.

Magnetic ordering in frustrated metallic systems depends in an
intricate way on the underlying electron Fermi surface. At small
filling factors, localized spins interact with each other through a
long-range oscillatory Ruderman-Kittel-Kasuya-Yosida (RKKY)
interaction mediated by the electrons. A Monte Carlo simulation of
pyrochlore spin-ice with RKKY interaction showed that the sign of
the effective Curie-Weiss constant as well as that of
nearest-neighbor coupling vary with the electron Fermi wavevector
\cite{ikeda08}. This in turn determines magnetic ordering at low
temperatures: the magnetic state of individual tetrahedra evolves
from the all-in all-out structure to the 2-in 2-out ferromagnetic
state with increasing electron density.

The geometry of the electron Fermi surface also plays an important
role in determining the magnetic instability of itinerant systems,
particularly at commensurate filling factors. A canonical example is
N\'eel ordering caused by perfect Fermi surface nesting that occurs
at a half-filled square-lattice Hubbard model. While the resulting
spin structure is collinear in bipartite square lattice, it was
shown that nesting effect on triangular lattice leads to a rare
noncoplanar magnetic order at filling factors $3/4$ and $1/4$
\cite{martin08,akagi10,kato10}. This chiral magnetic structure
induces a spontaneous quantum Hall effect in the absence of external
magnetic field. Although a long-range magnetic order cannot survive
thermal fluctuations in two spatial dimensions, the discrete
chirality order persists up to a finite temperature \cite{martin08}.

In this paper we demonstrate that the ground state of an isotropic
Kondo-lattice model on the pyrochlore lattice is magnetically
ordered with a noncoplanar spin structure and a definite chirality.
The conclusion also applies to Hubbard model on the pyrochlore
lattice at the mean-field level. The noncoplanar magnetic order
stems from a weak-coupling instability caused by perfect nesting of
Fermi `circles' at quarter filling; the quadrupled magnetic unit
cell contains 16 spins. In contrast to the triangular-lattice model
\cite{martin08}, this noncoplanar spin order does not support a
spontaneous Hall insulator because of a trivial Berry phase acquired
by the electrons when traversing a tetrahedron. The magnetic
structure itself, on the other and, has a definite handedness and is
characterized by a nonzero chiral order parameter.

It should be pointed out that although chiral magnetic orders have
been reported in metallic pyrochlore oxides such as
Pr$_2$Ir$_2$O$_7$ \cite{nakatsuji}, the non-coplanarity of spins in
these so~called spin-ice compounds are mainly created by a strong
uniaxial anisotropy. The electron-driven magnetic instability in
pyrochlore spin-ice at quarter filling will be briefly discussed at
the end of the paper.

We begin with the isotropic Kondo-lattice Hamiltonian on the
pyrochlore lattice:
\begin{eqnarray}
    \label{eq:H0}
    H &=& -t \sum_{\langle ij \rangle} \left(
    c^\dagger_{i\alpha}\,c^{\phantom{\dagger}}_{j\alpha} +
    \mbox{h.c.}\right) + J \sum_{\langle ij \rangle} \mathbf S_i\cdot\mathbf
    S_j \nonumber \\
    & & \! + \,\, K \sum_i \mathbf S_i\cdot c^\dagger_{i\alpha}\bm\sigma^{\phantom{\dagger}}_{\alpha\beta}
    c^{\phantom{\dagger}}_{i\beta}.
\end{eqnarray}
The first term describes electron hopping between nearest-neighbor
sites, $t$ is the hopping integral, and $c^{\dagger}_{i\alpha}$
creates an electron with spin $\alpha$ on site $i$. The second term
represents the superexchange interaction between neighboring
localized spins $\mathbf S_i$. The itinerant electrons interact with
the localized spins through an on-site exchange coupling $K$; the
vector $\bm\sigma_{\alpha\beta}$ denotes the three Pauli matrices.
We consider the classical limit of localized spins with magnitude
$|\mathbf S_i| = S \gg 1$. In this classical limit, the electron
spectrum is independent of the sign of $K$ and the eigenstates of
opposite signs are connected by a global gauge transformation
\cite{martin08}.

We first consider the tight-binding spectrum of electrons on the
pyrochlore lattice. In momentum space, the hopping Hamiltonian
[first term in Eq.~(\ref{eq:H0})] is expressed as $H_t =
\sum_{m,n}\sum_{\mathbf k,\alpha} \tau_{mn}(\mathbf k)\,
c^\dagger_{m \alpha}(\mathbf k)\, c_{n \alpha}(\mathbf k),$ where
$m,n$ are sublattice indices and the hopping matrix is
\begin{eqnarray}
    \label{eq:tau}
    \tau(\mathbf k) = -2\,t \left(\begin{array}{cccc}
    \mbox{\small 0} & c_{yz} & c_{zx} & c_{xy} \\
    c_{yz} & \mbox{\small 0} & \bar c_{xy} & \bar c_{zx} \\
    c_{zx} & \bar c_{xy} & \mbox{\small 0} & \bar c_{yz} \\
    c_{xy} & \bar c_{zx} & \bar c_{yz} & \mbox{\small 0}
    \end{array}\right).
\end{eqnarray}
Here we have introduced $c_{ab} = \cos\left[(k_a + k_b)/4\right]$
and $\bar c_{ab} = \cos\left[(k_a - k_b)/4\right]$ and set the
length of a cubic unit cell $a = 1$. This hopping matrix can be
exactly diagonalized, giving rise to a spectrum shown in
Fig.~\ref{fig:spectrum}(d). It consists of two degenerate flat bands
$\varepsilon^{\rm flat}_{\mathbf k} = 2 t$ and two dispersive bands:
\begin{eqnarray}
    \varepsilon^{\pm}_{\mathbf k} = -2\,t \left(1 \mp \sqrt{1+\mathcal{Q}_{\mathbf
    k}}\right),
\end{eqnarray}
where $\mathcal{Q}_{\mathbf k} = \cos\frac{k_x}{2}\cos\frac{k_y}{2}
+\cos\frac{k_y}{2}\cos\frac{k_z}{2}+\cos\frac{k_z}{2}\cos\frac{k_x}{2}.$
These two branches touch each other along the diagonal lines of the
square faces of the Brillouin zone [Fig.~\ref{fig:spectrum}(b)].

At quarter filling the lowest band $\varepsilon^-_{\mathbf k}$ is
completely filled. The corresponding Fermi `surface' is determined
by $\varepsilon^{+}_{\mathbf k} = \varepsilon^{-}_{\mathbf k}$ and
consists of diagonal lines $2\pi(x, 0, 1)$, and so on, on the square
faces of the zone boundary. The Fermi energy lies exactly at $E_F =
-2t$. Noting that $2\pi(x, 0, 1) \equiv 2\pi(1-x, 1, 0)$ module a
reciprocal lattice vector, these diagonal lines are topologically
equivalent to three closed loops, or circles, intersecting with each
other at the X points of the Brillouin zone
[Fig.~\ref{fig:spectrum}(b--c)]. In the vicinity of these Fermi
lines, the spectrum has the form $\varepsilon^{\pm}_{\mathbf k}
\approx E_F \pm C_{\varphi}\times |\mathbf k_{\perp}|$, where
$\mathbf k_{\perp}$ is the perpendicular component of electron
momentum.

The reduction of Fermi surface to Fermi lines at quarter filling
also makes it possible to realize perfect nesting with a few
wavevectors. Indeed, the three Fermi circles can be perfectly nested
by wavevectors $\mathbf Q_1 = 2\pi(1,0,0)$, $\mathbf Q_2 =
2\pi(0,1,0)$, and $\mathbf Q_3 = 2\pi(0,0,1)$, respectively. For
example, states on the two Fermi lines $2\pi(1,0,z)$ and
$2\pi(0,1,z)$, both belonging to the blue Fermi circle in
Fig.~\ref{fig:spectrum}(c), differ by $\mathbf Q_3$ up to a
reciprocal lattice vector. The Van Hove singularity of the density
of states at the Fermi level gives rise to a logarithmically
divergent Lindhard susceptibility: $\chi(\mathbf q) \sim
-\log|\mathbf q - \mathbf Q_{\eta}|$, indicating a weak-coupling
instability caused by the nesting effect.

To investigate possible magnetic ordering caused by perfect nesting
of the Fermi lines, we consider spin orders which can be expanded as
\begin{eqnarray}
    \label{eq:spin}
    \mathbf S_i = \mathbf S_m(\mathbf r) = \sum_{\eta=0}^3 \mathbf
    M_{m,\eta}\,e^{i\mathbf Q_{\eta}\cdot\mathbf r},
\end{eqnarray}
where $m$ is the sublattice index and $\mathbf r$ denotes the
position of the Bravais (fcc) lattice point. In order to
accommodate, e.g. ferromagnetic ordering, we have included a
$\mathbf Q_0 = 0$ component in the above expansion. The unit cell of
this multiple-$\mathbf Q$ magnetic order is extended to the
conventional cubic unit cell which contains 16 inequivalent spins.

Next we include the on-site Hund's exchange which couples electron
with momentum $\mathbf k$ to that with $\mathbf k + \mathbf
Q_{\eta}$. After Fourier transformation, the on-site coupling term
has the form
\begin{eqnarray}
    H_K = \sum_{\xi,\eta}\sum_{\alpha,\beta}\sum_{m,\mathbf
    k}\!'\, c^\dagger_{m \alpha}(\mathbf k\!+\!\mathbf Q_{\xi})\,\,
    \mathcal{M}^{(m)}_{\xi\alpha, \eta\beta}\,\,
    c^{\phantom{\dagger}}_{m \beta}(\mathbf k\!+\!\mathbf Q_{\eta}).
    \,\,\,
\end{eqnarray}
Here the prime indicates that the momentum summation is restricted
to the reduced Brillouin zone, and the coupling coefficients are
given by
\begin{eqnarray}
    \mathcal{M}^{(m)}_{\xi\alpha,\eta\beta} = K \,
    \Delta_{\xi\eta,\,\zeta}\, (\mathbf
    M_{m,\zeta}\cdot\bm\sigma_{\alpha\beta})
    \,e^{i(\mathbf Q_\xi -
    \mathbf Q_\eta)\cdot\mathbf d_m},
\end{eqnarray}
where $\mathbf d_m$ denotes basis vectors of the pyrochlore lattice.
Momentum conservation of the coupling is encoded in the symbol
$\Delta_{\xi\eta,\,\zeta}$ which is symmetric with respect to the
first two indices. The nonzero components include $\Delta_{01,1} =
\Delta_{02,2} = \Delta_{03,3} = \Delta_{23,1} = \Delta_{31,2} =
\Delta_{12,3} = 1$.

\begin{figure}
\includegraphics[width=0.96\columnwidth]{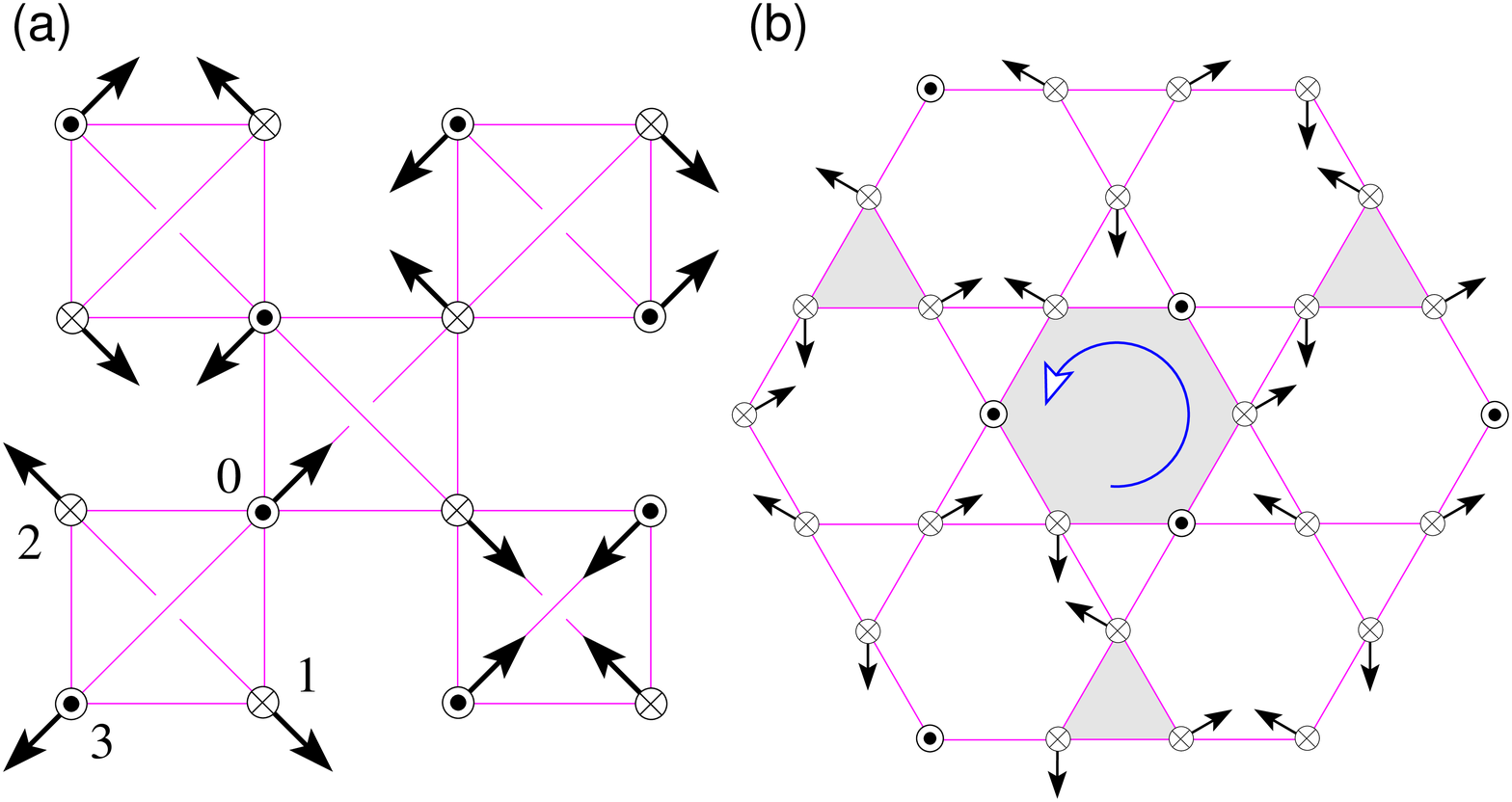}
\caption{\label{fig:m-order} (a) A unit cell of the noncoplanar
magnetic order viewed from the $c$ axis. Here we choose $\hat\mathbf
n_1 = \hat\mathbf a$, $\hat\mathbf n_2 = \hat\mathbf b$, and
$\hat\mathbf n_3 = \hat\mathbf c$ in Eq.~(\ref{eq:m-order}). (b)
Chiral structure of the same magnetic order projected onto a $(111)$
kagome plane. The $\odot$ and $\otimes$ symbols denote spin
component coming out of and into the plane, respectively. Same
handedness (right-handed in this case) is observed in other kagome
planes.}
\end{figure}

For a given set of Fourier components $\mathbf M_{m,\eta}$,
diagonalization of the electron Hamiltonian $H_t + H_K$ gives a
total of 32 energy bands. Due to perfect nesting of Fermi lines, the
first 8 bands are expected to be separated from other high energy
states by an energy gap. At quarter filling, the ground-state energy
of electrons $E_e$ is obtained by filling the lowest eight bands.
However, the determination of the minimum-energy state is not
straightforward because of the large number of variables required to
describe the spin structure. Indeed, even restricting to magnetic
orders which can be expressed in the form of Eq.~(\ref{eq:spin}),
one still needs 32 variables to specify the 16 classical spins in a
quadrupled unit cell. Here we employ the simulated annealing
algorithm to minimize the energy functional $E_e\left(\mathbf
M_{m,\eta}\right)$. For Hund's coupling below a critical value $K_c
\approx 4.32\,t$ and $J = 0$, we obtained the same noncoplanar
magnetic order shown in Fig.~\ref{fig:m-order} using different
initial configurations and rates of decreasing the effective
temperature. For $K > K_c$ ferromagnetic order with $\mathbf S_i =
S\,\hat\mathbf n$ takes over and becomes the ground state.

To characterize the noncoplanar spin structure, we introduce three
order parameters $\mathbf L_1 = (\mathbf S_0 + \mathbf S_1 - \mathbf
S_2 - \mathbf S_3)/4S$, and so on  \cite{chern06}, which measure the
staggered magnetizations of a tetrahedron; the spin subscript
indicates the corresponding sublattice. A unit cell of the magnetic
order is shown in Fig.~\ref{fig:m-order}(a), where the magnetic
states of type-I tetrahedra are described by order parameters
\begin{eqnarray}
    \label{eq:m-order}
    \mathbf L_1 = \frac{\hat\mathbf n_1}{\sqrt{3}}\,e^{i\mathbf
    Q_{3}\cdot\mathbf r},\,\,\,\, \mathbf L_2 = \frac{\hat\mathbf
    n_2}{\sqrt{3}}\,e^{i\mathbf Q_{1}\cdot\mathbf r}, \,\,\,\, \mathbf
    L_3 = \frac{\hat\mathbf n_3}{\sqrt{3}}\,e^{i\mathbf
    Q_{2}\cdot\mathbf r}.\,\,
\end{eqnarray}
Here $\hat\mathbf n_r$'s are three mutually orthogonal unit vectors.
Remarkably, the noncoplanar magnetic order is a simultaneous ground
state of the inter-site exchange interaction. To see this, we note
that the superexchange term in Eq.~(\ref{eq:H0}) can be recast into
$H_J = (J/2)\sum_{\boxtimes} \left|\mathbf
M_{\boxtimes}\right|^{2}$, where $\mathbf M_{\boxtimes} =
\sum_{m=0}^3 \mathbf S_m$ is the total magnetization of a
tetrahedron. For the magnetic order shown in Fig.~\ref{fig:m-order},
the total spin of every tetrahedron, including both types, vanishes
identically, hence minimizing the exchange energy. This result also
indicates that the ferromagnetic order favored by a large Hund's
coupling $K$ is suppressed by the antiferromagnetic inter-site
exchange.

\begin{figure}
\includegraphics[width=0.65\columnwidth]{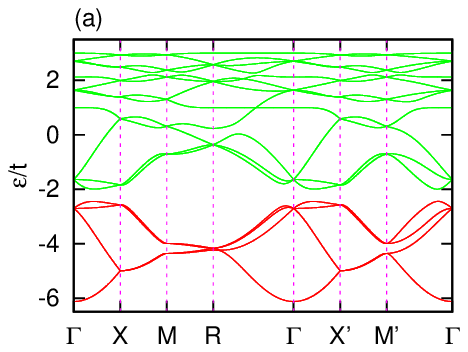}
\includegraphics[width=0.33\columnwidth]{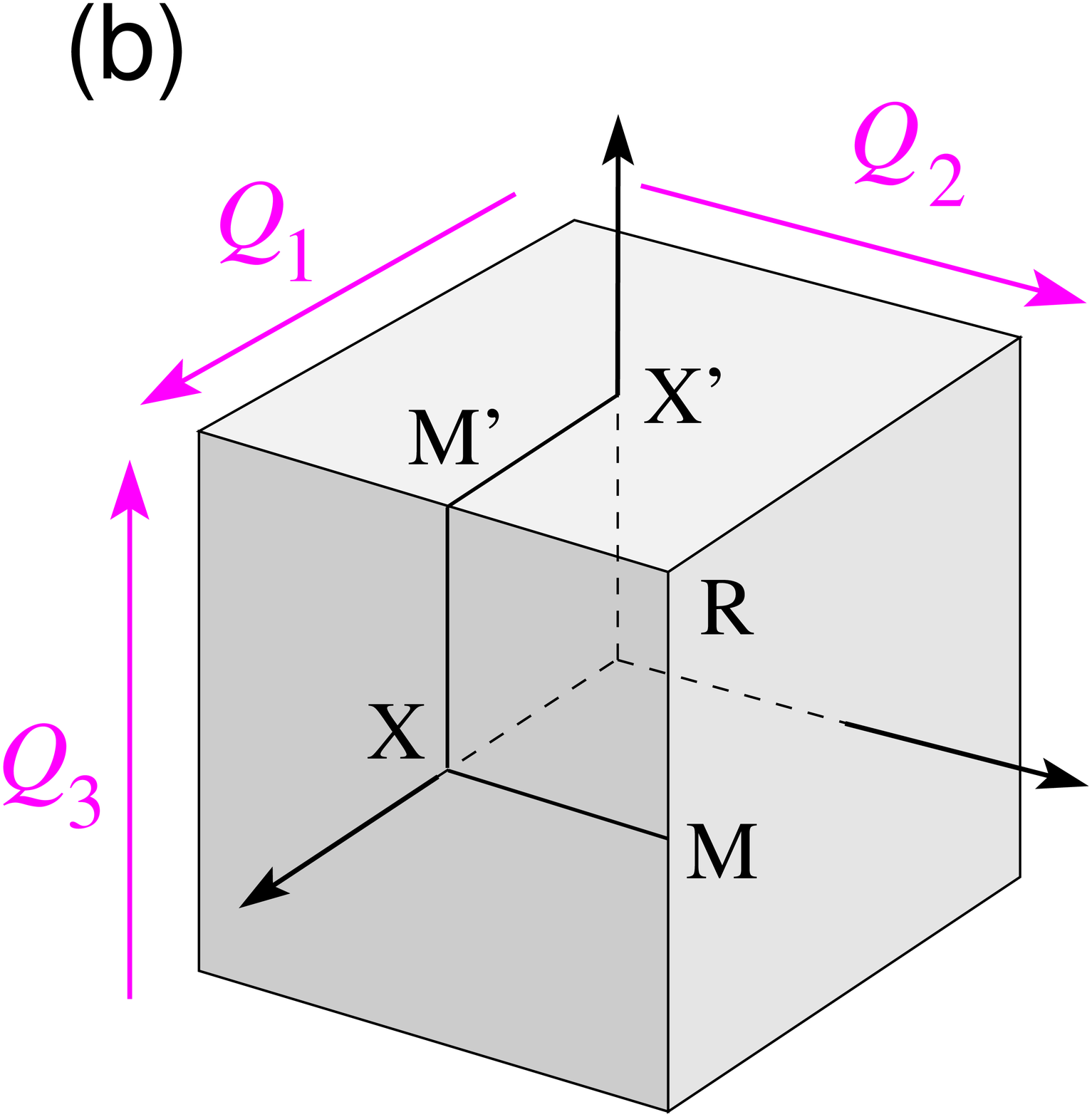}
\caption{\label{fig:ek} (a) Electron band structure corresponding to
the noncoplanar magnetic order shown in Fig.~\ref{fig:m-order}. The
Hund's coupling is $K = t$ and superexchange constant $J = 0$. (b)
The reduced Brillouin zone with high-symmetry points and lines.}
\end{figure}

Fig.~\ref{fig:ek}(a) shows the electron band structure corresponding
to the noncoplanar magnetic order. The spectrum is at least doubly
degenerate thanks to a combined lattice translation and spin
rotation symmetry. Due to the nesting effect, a charge gap opens
between the lower four pairs of bands and other high energy
branches. It is known that the electron experiences a fictitious
magnetic field proportional to the spin chirality $\chi_{ijk} =
\mathbf S_i\cdot\mathbf S_j\times \mathbf S_k$ \cite{nakatsuji} when
hopping around a triangular loop. Although $\chi_{ijk}$ is nonzero
in the noncoplanar structure, the contributions from the four
triangular faces of a tetrahedron cancel each other due to a
monopole-like spin configuration. The noncoplanar magnetic order
thus does not exhibit a spontaneous quantum Hall effect.

\begin{figure} [t]
\includegraphics[width=0.95\columnwidth]{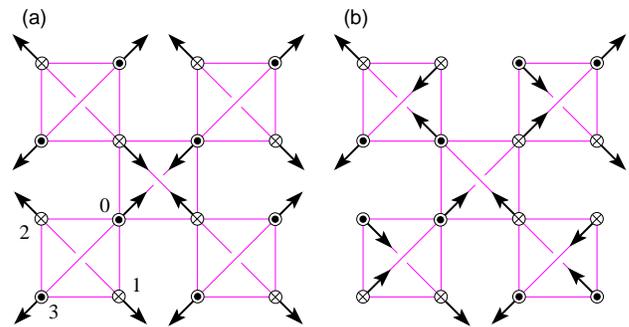}
\caption{\label{fig:m-ice} (a) All-in all-out structure. (b) 2-in
2-out magnetic order with wavevecotr $\mathbf Q_3 = 2\pi(0,0,1)$.
The $\odot$ and $\otimes$ symbols again denote spin component coming
out of and into the plane, respectively.}
\end{figure}

The magnetic structure itself, nonetheless, breaks the chiral
symmetry. This is best illustrated by projecting the triple-$\mathbf
Q$ magnetic order onto a $(111)$ kagome plane which shows hexagons
and triangles with a definite handedness
[Fig.~\ref{fig:m-order}(b)]. More specifically, we introduce a
scalar chirality for a tetrahedron in the antiferromagnetic state:
\begin{eqnarray}
    \chi_{\boxtimes} = 3\sqrt{3}\,\,\mathbf L_1 \cdot \mathbf L_2 \times
    \mathbf L_3.
\end{eqnarray}
The noncoplanar magnetic order shown in Fig.~\ref{fig:m-order} has
the maximum chirality $\langle \chi_{\boxtimes} \rangle = + 1$,
where the average is over both types of tetrahedra. This is to be
contrasted with the so called all-in all-out structure
[Fig.~\ref{fig:m-ice}(a)] in which the two types of tetrahedra have
opposite chiralities $\langle \chi_{\rm I} \rangle = - \langle
\chi_{\rm II} \rangle$; the average chirality is thus zero.

The model described by Eq.~(\ref{eq:H0}) with a negative $K$ was
recently investigated in Ref.~\onlinecite{motome10} using Monte
Carlo simulations. The authors also considered the special case of
quarter filling. However, since the simulations were performed
assuming a large Hund's coupling, the noncoplanar magnetic order
shown in Fig.~\ref{fig:m-order} was not observed in their
simulations. Although the fine structure of the electron Fermi
surface is absent in small finite systems, the noncoplanar magnetic
order could be found in Monte Carlo simulations at small Hund's
coupling using techniques such as twisted boundary conditions
\cite{motome10}.

We now turn to the spin-ice compounds where the rotational symmetry
is explicitly broken by a strong easy-axis anisotropy $H_D = -D
\sum_i (\mathbf S_i\cdot\hat\mathbf e_m)^2$. The effective degrees
of freedom become discrete Ising variables as spins are forced to
point along the local $\langle 111 \rangle$ directions. We start
with the $J = 0$ limit. At filling factor $\nu = 1/4$, simulated
annealing yields a ground state with the all-in all-out magnetic
order shown in Fig.~\ref{fig:m-ice}(a). The fact that every
tetrahedron has total spin $\mathbf M_{\boxtimes} = 0$ makes the
all-in all-out structure a simultaneous ground state of the
antiferromagnetic inter-site couplings.

In spin-ice compounds, however, the nearest-neighbor exchange has a
ferromagnetic sign, i.e. $J = -|J|$. Assuming $D \gg |J| \gg t$,
$K$, numerical minimization gives a single-$\mathbf Q$ 2-in 2-out
structure [Fig.~\ref{fig:m-ice}(b)]. The magnetizations of type-I
and type-II tetrahedra point along $a$ and $b$ axes, respectively,
with their sign alternating between adjacent $ab$ planes. There is a
total of six degenerate single-$\mathbf Q$ states related by time
reversal and lattice translation symmetries.

Both magnetic orders shown in Fig.~\ref{fig:m-ice} were obtained in
a recent Monte Carlo study on a metallic pyrochlore spin ice
\cite{ikeda08}. Instead of tackling the Kondo-lattice model
directly, the simulations were performed assuming a long-range RKKY
interaction between localized spins. Neglecting the superexchange
$J$, their simulations yielded a 2-in 2-out structure at quarter
filling \cite{ikeda08}, which is in disagreement with our result
(the all-in all-out state). This discrepancy could be attributed to
the fact that the RKKY interaction is derived by integrating out the
electrons with a spherical Fermi surface. The approach thus neglects
the subtle effect of Fermi surface nesting.

In summary, we have studied a model of itinerant electrons
interacting with localized classical spins on the pyrochlore
lattice. At quarter filling, we showed that a noncoplanar magnetic
order emerges as a ground state of the rotationally invariant
Hamiltonian. The magnetic structure characerized by a nozero order
parameter $\chi_{\boxtimes}$ also breaks the chiral symmetry. Since
the chiral transition needs not coincide with the magnetic ordering
from the symmetry viewpoint, a chiral phase with $\langle
\chi_{\boxtimes} \rangle \neq 0$ but no spin order $\langle \mathbf
L_i \rangle = 0$ could occur at finite temperatures. A similar case
was recently observed in spin ice compound Pr$_2$Ir$_2$O$_7$
\cite{machida}. Future studies, especially large-scale Monte Carlo
simulations, could demonstrate the noncoplanar magnetic structure
numerically and explore the possibility of the partially ordered
chiral phase. Finally, we showed that when the localized spins are
subject to strong uniaxial anisotropy, the on-site Hund's coupling
favors a uniform all-in all-out spin structure, whereas the
single-$\mathbf Q$ 2-in 2-out magnetic order is stabilized by the
inter-site ferromagnetic exchange interaction.

{\em Acknowledgment.} I am grateful to Y. Kato, I. Martin, Y.
Motome, and N. Perkins for stimulating discussions, and especially
to C.D. Batista for pointing out the logarithmic divergence of
susceptibility and various invaluable comments. The author also
acknowleges the support of ICAM and NSF Grant DMR-0844115.

\end{document}